# Causal Analysis of Generic Time Series Data Applied for Market Prediction

Anton Kolonin[1,2,3][0000-0003-4180-2870], Ali Raheman[1], Mukul Vishwas[1][0000-0002-8824-1954], Ikram Ansari[1][0000-0002-9091-6674], Juan Pinzon[1][0000-0001-9062-683X], and Alice Ho[1][0000-0003-1339-3969]

[1] Autonio Foundation Ltd., Bristol, UK
[2] SingularityNET Foundation, Amsterdam, Netherlands
[3] Novosibirsk State University, Novosibirsk, Russian Federation
`{akolonin, ali.raheman}@gmail.com`

**Abstract.** We explore the applicability of the causal analysis based on temporally shifted (lagged) Pearson correlation applied to diverse time series of different natures in context of the problem of financial market prediction. Theoretical discussion is followed by description of the practical approach for specific environment of time series data with diverse nature and sparsity, as applied for environments of financial markets. The data involves various financial metrics computable from raw market data such as real-time trades and snapshots of the limit order book as well as metrics determined upon social media news streams such as sentiment and different cognitive distortions. The approach is backed up with presentation of algorithmic framework for data acquisition and analysis, concluded with experimental results, and summary pointing out at the possibility to discriminate causal connections between different sorts of real field market data with further discussion on present issues and possible directions of the following work.

**Keywords:** Causality, Causal Analysis, Correlation, Financial Market, Time Series.

## 1 Introduction

### 1.1 Background for this work

The motivation of this work is to figure out a suitable general purpose algorithmic framework capable of figuring out causal connections across diverse time series data from different sources, including sparse and unreliable ones. The motivation is supported by our further work on the generic architecture for active portfolio management [1] employed by automated adaptive trading and market making agents [2] which need to be capable to do predictions in respect to future market dynamics relying on diverse temporal streams of data. This includes market data, social and online media news, as well as so-called "on-chain" data computed from transactional activities on public financial ecosystems such as blockchains.

While we understand that the operations being performed by a hypothetical completely autonomous trading or market making agent might be considered as a narrow artificial general intelligence (Narrow AGI), we want to have the operational environment of it to gain as much reach as possible, maximizing its capabilities for intelligent decision making based on wide range of information sources, including market data and technical indicators from different exchanges, fundamental and "on-chain" data, and sentiment and emotional data from online and social media sources. That is why in this work we explore the possibility of causal analytics for market prediction purposes for as much information as possible given rather specific business case of the Bitcoin price prediction on Binance exchange for BTC/USDT pair referring to Tether USD stable coin.

### 1.2 Overview of the field

The fundamental background for probabilistic causal analytics can be found in [3] with application of predictive causal analytics to financial markets discussed in [4]. The recent study of causal analytics applied to time series data is covered in [5]. Application of sentiment analysis in respect to causal analysis of sentiment data and market volatility on its basis is presented in [6]. The variety of features, metrics, and parameters then can be derived from the market data, including the structure of the limit order book (LOB) snapshots is covered in [7] and [8]. Finally, the very latest study discovers the connection between patterns in political and economic history with so-called "cognitive behavioral schemata" (CBS) patterns traditionally used in psychotherapy [9]. All the mentioned studies have been accounted, extended and tailored to the specific problem in hands as discussed further.

## 2 Practical approach

### 2.1 Data acquisition

Given the practical objective of our work is providing operations on crypto exchanges such as Binance and the crypto finance is a domain being actively discussed on social media channels such as Twitter and Reddit, we have tried to collect as much as possible data from both kind of sources.

**Market Data.** In particular, the present data acquisition framework streams the live market data from Binance exchange, including both raw trades and snapshots of the LOB at different sampling rates or granularity periods including 1 day, 1 hour, 1 minute, and 1 second. Both sorts of the mentioned "raw" data were used to compute the "pre-processed" data such as extended open-high-low-close-volume (OHLCV) frames, including volumes and counts of "buy" and "sell" (from the regular trader perspective) traded, average prices for "buy" and "sell" trades, including regular averages as well as weighted averages using both base and quote currency for the averaging weights. All the counts, volumes, and average prices for "buy" and "sell" are used to compute "imbalance" metrics indicating the skew of the distribution towards either "buy" or "sell". That is, we have substantially extended the scope of features used in [7]. The use of LOB data has been rendered useful in [8], so we include more metrics shaping the distribution of the orders such as minimum "ask" and maximum "bid" prices, average "ask" and "bid" prices with the averages weighted by order volumes, "spread" and average "spreads" between different sorts of the "ask" and "bid", order volumes on each of the order book sides and all sorts of imbalances on these "ask"/"bid" prices and volumes. The overall scope of the market data for the BTC/USDT pair discussed in this work was covering almost 1.5 years from August 2020 till December 2021.

The "pre-processed" data described above have been normalized in different alternative ways in order to turn them into stationary state in range between [-1.0,+1.0]. All the data was differentiated so the derivatives were computed on basis of the raw data. Next, the differentiated data and the raw data has been turned too non-negative logarithmic scale using operation *log10(1+x)*. Finally, from this point, both differentiated and non-differentiated, logarithmic and non-logarithmic data has been normalized using operation *x/max(abs(x))* to ensure the range [-1.0,+1.0] regardless of the metric sign. It worth noticing that at the earlier phase of the work different normalization schemes were applied, however it was found that some of the metrics perform better in the representations other than expected, so eventually we have decided to apply all sorts of normalization to every metric at the cost of increased number of times series involved in the analysis on the following phase.

**Media Data.** Two kinds of metrics were derived from the online social media data: public posts from about 80 channels on Twitter and Reddit relevant to crypto market for six months starting July 2021. First, it was the conventional sentiment as presented in [6], computed as described below. Second, it was the "cognitive behavioral schemata" (CBS) patterns evaluated according to [9]. The overall volume of the media content was exceeding 100,000 posts across all channels.

The sentiment metrics were computed with help of Aigents®, which is "interpretable" model based on "n-grams," available as part of https://github.com/aigents/aigents-java distribution and written in Java, which comes with "out-of-the-box" vocabularies for n-grams associated with positive and negative sentiment. It has over 8,200 negative and over 3,800 positive n-grams and returns the overall sentiment/polarity of the text based on the frequencies of occurrences of the reference n-grams in the text along with independent positive and negative sentiment metrics. One of the specifics of the model is implementation of the "priority on order" principle as discussed in [10]. In the Aigents®-specific implementation it means precedence given for n-grams with higher "n", so whenever any n-gram is matched, all matches of any other n-grams being parts of the former n-gram are disregarded. For instance, if tetragram ["not","a","bad","thing"] is matched, then both bigram ["bad","thing"] and unigram ["bad"] are disregarded and discounted. Similarly, matching bigram ["no", "good"] disregards and discounts both constituent unigrams ["no'] and ["good"]. In addition to that, the model has an option to provide logarithmic scaling of the counted frequencies and our studies have revealed that by enabling this option it provides better performance. The model provides four basic sentiment metrics, so that, instead of addressing the sentiment analysis problem as a plain classification ('Positive' vs. 'Negative' vs. 'Neutral'), we were treating it as a multinomial classification problem in four independent dimensions corresponding to the individual metrics discussed below:

*sentiment (sen)* – overall or compound sentiment/polarity in range [-1.0,+1.0], so its value can be either negative or positive;

*positive (pos)* – canonical positive sentiment assessment in range [0.0,+1.0], so its value can be only positive;

*negative (neg)* – canonical negative sentiment assessment in range [-1.0,0.0], so its value can be only negative;

*contradictive (con)* – mutual **contradictiveness** of the positive and negative assessments computed as *SQRT(positive * ABS(negative))*.

All of the media metrics computed on basis of individual posts were aggregated as mean/average per channel across all channels on either daily or hourly basis and the aggregated mean values were in turn normalized using operation *x/max(abs(x))* to ensure the range [-1.0,+1.0].

## 2.2 Analytical framework

Since the practical and goal of the study was the prediction of the market price, our causal analytical framework was considering the price movement as a target "effect" and all the other metrics as a potential "causes". While the earlier work [7] refers to stationary function of "log-return" as a target, we were dealing with price difference (price derivative, PD) after finding that fundamental nature of results presented further does not depend on that choice while use of PD had turned to be handier in practical applications. That is, the PD was considered as the "effect".

The conceptual causal frameworks [3] and [4] justifying our studies has turned difficult to implement literally due to the lack of clearly identifiable "events" in the time series data, even assuming the data is represented by stationary functions in the range [-1.0,+1.0]. It was tempting to consider determination of events such as "price goes up", or "there is positive sentiment" but it was clear that it could be done on basis on some thresholds which would be either subjective or become a source of extra errors and uncertainties or both. On the other hand, the formal assessment of probability of such "event" would become another problem. One would suggest using the values of the "effect" and "cause" metric functions in the range [-1.0,+1.0] as probabilities but it could be ambiguous either because of diversity of scaling factors for individual metrics being forcefully aligned to the same normalization range. After all, we ended up following the approach of temporally lagged/shifted correlation analysis applied to time series data as described in [5] and [6]. That is, we were considering the **causation** as the **preceding correlation**, or correlation of the "cause" function with the "effect" function shifted back to certain lag on the temporal axis.

Given the rich data that we had, we were performing the causal analysis in three dimensional space, with time *t* being the first dimension, channel *c* being the second dimension and the metric *m* being the third one. The channel might be either actual Twitter or Reddit channel used to derive the media metric or some source of the market data (such as Binance) or "on-chain" data (such as Bitcoin) or third-party sources

(such as Santiment or Glassnode services discussed further). The metrics would be specific in respect to the channel. The results presented in the next section were derived on basis of 80 media channels with 21 metrics in each and 111 metrics in single market data channel corresponding to BTC/USDT trading pair on Binance exchange, so 1791 potential "causes" were explored in total, in different time sampling rates such as day, hour, minute and second.

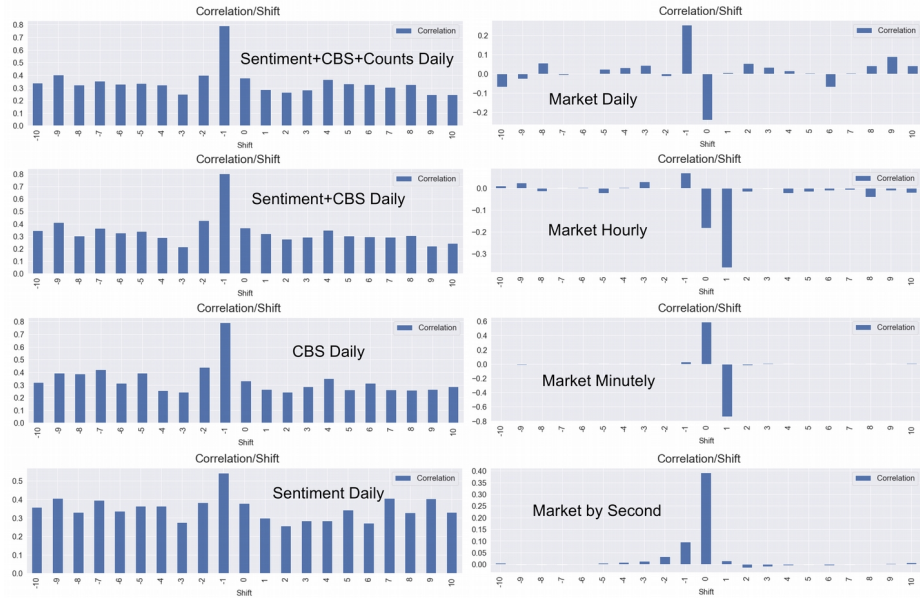

**Fig. 1.** Search for preceding correlation between the effect PD and synthetic additive cause indicator (SACI) on different temporal horizons measured as shift of the effect function back in time (negative shift to the left) or forward in time (positive shift to the right). The left bar charts present respective Pearson correlation (PC) of the shifted price with SACI assembled using different sets of media metrics across all Twitter and Reddit channels (top to down): sentiment with CBS with word count and post count, sentiment with CBS, CBS only, sentiment only – everything on daily basis. The right bar charts present PC of the shifted price with SACI assembled using all market metrics based on Binance data (top to down): on daily basis, on hourly basis, on minutely basis, on per-second basis.

The key study was the process of finding what we called synthetic additive cause indicator (SACI) relying on the whole scope of source metrics being treated as a hypothetical causes. The probabilistic logic treats addition as logical disjunction and multiplication as logical conjunction. In this work we were exploring only the disjunctive version of it, so the assembly of the integrative SACI was involving addition of the perspective causes in order to maximize the correlation with the effect at a particular target shift/lag. See the discussion on the SACI performance presented on Fig.1 in the following section.

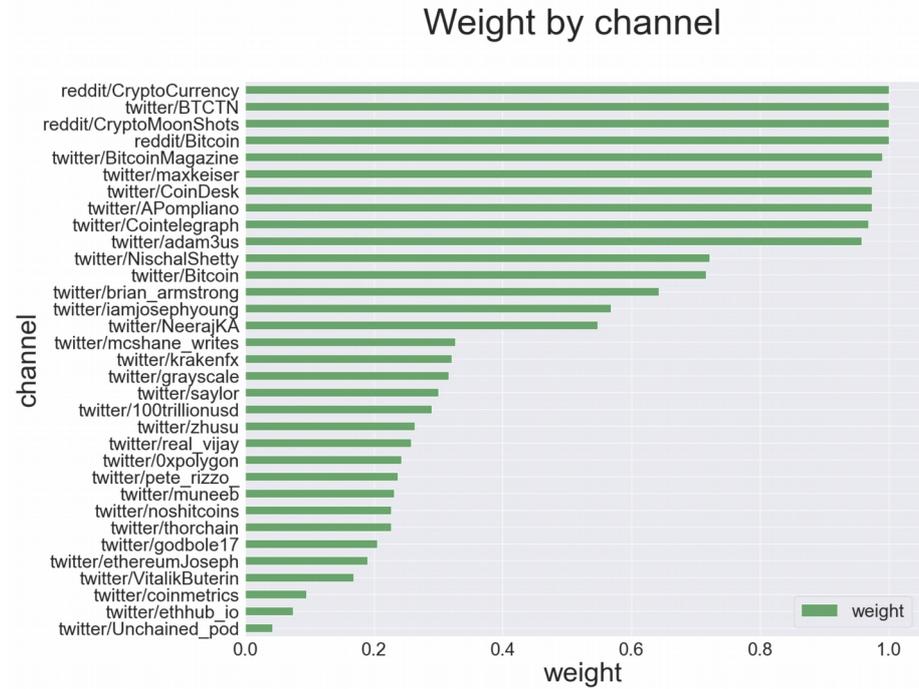

**Fig. 2.** Presentation of the "representability weight" of each of the media channels involved in the search for the SACI on daily basis, the study is showing that account on the weight while assembling the SACI improves the fitting of the cause-effect correlation on the training data set. The top channels with the weight equal to 1.0 have posts on every day while channels at the bottom with the weight close to zero have just few posts per month/week.

The temporal causation study was run evaluating different time shifts/lags in days [-10,+10] computing mutual Pearson correlation (PC) between each of the potential causes and the price difference and retaining the "correlation weights" of the computed value $P(l,c,m)$ for every time lag $l$, news channel $c$, and metric $m$. Also, the channels $c$ were optionally weighted with the "representability weight" as $W(c)$ according to the percentage of days (or hours) with news present on such time intervals, as shown on Fig.2. Then, for every lag $l$, the compound SACI metric time series $Y(l,d) = \Sigma X(c,m,d)*P(l,c,m)*W(c)$ for every day $d$ have been built from the original raw metrics $X(c,m,d)$. The compound SACI metric building process was implemented starting from channels with the highest $W(c)$ and $P(l,c,m)$ adding ingredients up to $Y(l,d)$ incrementally, as long as the correlation between the target price difference function and the current content of summed up $Y(l,d)$ series for given time lag $l$ keeps increasing.

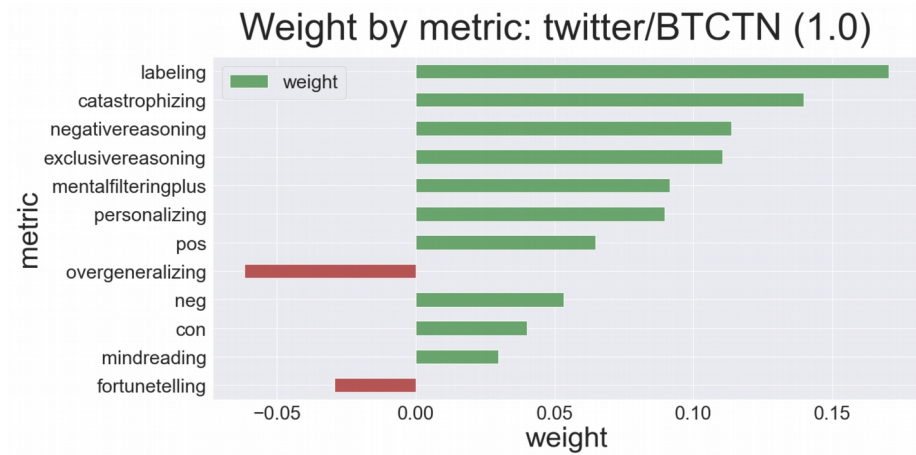

**Fig. 3.** Study of the "correlation weights" for one single media channel being part of the model as preceding PC at shift/lag one day before an "effect", showing that the most impactful on the price change appear to be the specific cognitive distortions called "labeling" and "catastrophizing". The positive weight means that either increase of these distortions is preceding price increase, which appears more reasonable due to the expectedly speculative nature of the crypto market or decrease of them is preceding the price decline. It is noticeable that negative weight of the "overgeneralizing" distortion apparently corresponds to the opposite – either increase of this distortion is preceding price decline, which appears more reasonable, or decrease of it is preceding the price increase. In accordance with the findings discussed regarding the Fig.1, the sentiment metrics (*pos*, *con* and *neg*) appears much less impactful in respect to the price change, rendering the high degree of **contradictiveness** in respect to each other.

## 3     Experimental results

The **causal connectivity as** a preceding correlation has been studied on the full scope of the media and market data described above with major results presented on Fig.1. It is clearly seen that ability to build the well-correlated SACI from media data at the point one day before the anticipated "effect" is dominating all other time lags/shifts so we can with a greater certainty state that some combination of the metrics represented by the "model" of the SACI is having the causal connection with the target price change. In turn, the "model" of the SACI represented by the number of the channels and metrics involved in it along with their "correlation weights" and "representability weight" are determining the fine-grained causal structure of it discussed further. It is also seen that sentiment doesn't have significant impact on the causation alone (PC = 0.56), the involvement of word and news counts make the results a little bit worse (PC going down from 0.8 to 0.78), the CBS alone provides PC = 0.79 and CBS with sentiment together bring it to the maximum (PC = 0.8).

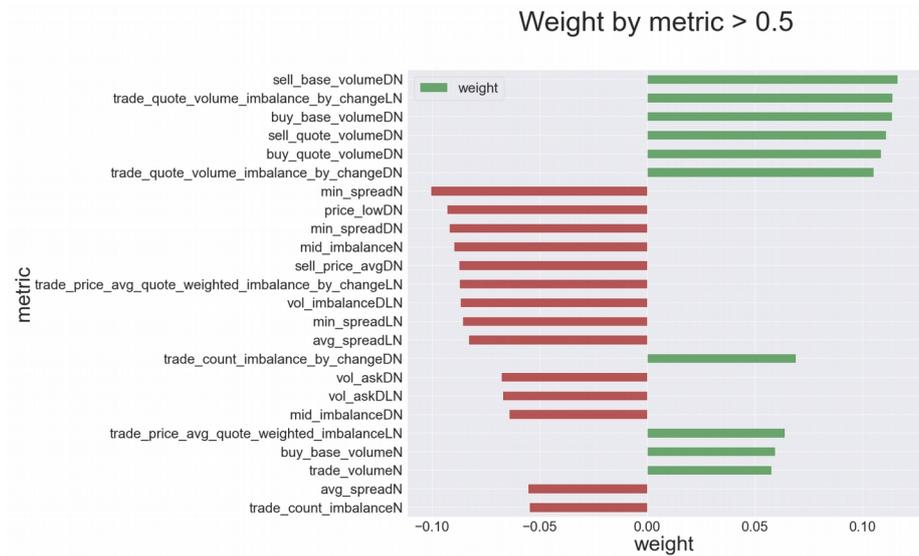

**Fig. 4.** A Study of the "correlation weights" for the market metrics, showing the less expression of the preceding PC at one day shift/lag before an "effect". The capital letters in feature suffixes correspond to involved "pre-processing" applied in left-to-right order: D – derivative or differential, L – decimal logarithm, N – normalization to range [-1.0,+1.0]. It is clearly seen that the most impactful features (PC > 0.1) appear to be the volumes of "sell" and "buy" trades and the imbalance between them denominated by the price change magnitude at the moment (*trade_quote_volume_imbalance_by_change*). Also, in accordance with discussion on Fig.1, the PC assessments for market metrics are substantially less impressive than for media metrics.

While the sentiment metrics have appeared promising thus far, the market metrics have turned to be substantially less inspiring. The daily study for market metrics on Fig.1 do render promising correlation of the SACI one day before the "effect", however the low PC = 0.25 at this shift is much less than in case of using media metric and there is much more expressive correlation coinciding with the "effect" at zero shift/lag with almost the same PC value (negative in this case). Moreover, the studies for hourly, minutely and by-second sampling rates do not render noticeable preceding correlations at all.

The extra data involved for this kind of analysis were the pre-syndicated media, market and related data by third-party providers. Specifically, we explored the daily and hourly data feeds from Santiment (https://api.santiment.net) to check for sentiment and on-chain metrics both and Glassnode (https://glassnode.com) to check for on-chain data only. The on-chain metrics are indicators derived from different sorts of transnational activity on blockchain such as Bitcoin. The period for study was taken the same as for the social media feeds discussed above – nearly half year starting July 2021. While working with the Santiment API, we looked at various channels like Telegram, Reddit, Twitter and Bitcointalk with each of the channels supplied with negative and positive sentiment metrics provided by Santiment service. We also considered non-sentiment metrics from it, like circulating supply, active addresses, and GitHub activity available on the platform. But we could not identify any metrics that could impact the price prediction. The on-chain metrics used from Glassnode involved active address count, transactions count, transactions rate, blockchain count, blockchain height, grayscale holdings, to name a few. The study has shown many metrics having positive Pearson correlation synchronously, at the same day or hour with the price change (measured as price derivative or "log return"), yet no one was showing expressed causal connection with the price change in terms of preceding correlation on the shifted time series, so no further studies has been done on this data. Notably, the synchronous and preceding correlations were higher on daily data but much weaker on hourly data as it was found for other data discussed above.

The **causal structure** of the additive ingredients of the SACI rendering the highest preceding PC scores at one day lag/shift before the price change "effect" was done as shown on the Fig.3 and Fig.4 for media and market metrics, respectively. It shows confirmation on the greater potential applicability of the media metrics over the market metrics and use of the cognitive distortions over the sentiment, specifically.

### 3.1 Practical applications

The results presented above have been tried to get applied for price prediction of the BTC/USDT (Bitcoin to USD Tether) trading pair on Binance exchange. The objective has been set to hit two targets. First, we were looking to exceed the baseline provided by "prediction" made just by copying the "last known price" (LKP) and approach the "prediction" made by looking up the "future known price" (FKP) in historical test data. The performance has been evaluated on basis of both Mean Average Percentage Error (MAPE) and Directional Accuracy (DA). The data used for experiments were the same as discussed above. Second, we were using our backtesting framework [1,2] to use obtained predictions by the market making bots according to their strategies.

So far, in order to accomplish the goal, we tried classical Machine Learning algorithms such as Linear Regression, Ridge, Lasso regressions and Elastic Net among others without any clear success to outperform the LKP baseline. We experimented with the Long Short-Term Memory (LSTM) artificial recurrent neural network architecture. We did extensive feature engineering ending up with 53 input features for our LSTM model, these features were from OHLCV and Limit Order Book data plus calculating some basic Technical Analysis indicators such as RSI, MACD and Moving Averages among others. Normalizing our feature set was required to transform all features into homogeneous values, MinMax scaler proved to provide better results. We tested our model with different amounts of training data, historical intervals and different hyper-parameters for different data intervals, 1 minute, 1 hour, 4 hours and 1 day. Only when we did an 'Ensemble' of 5 of our tested LSTM configurations predictions we managed to outperform the Last Known Price baseline when predicting a couple of days of June 2021. Unfortunately, these results did not translate when predicting full month periods or in our backtesting framework across most of the months over the years 2020-2021. Our LSTM Ensemble model proved to be very susceptible to market conditions, where bearish market conditions in May 2021 made possible some surprisingly good results in our backtesting framework, so the market making agents using the predictions were getting substantially larger revenues than the agents not using the predictions, even though MAPE and DA of these predictions was not exceeding the LKP as an average.

## 4     Conclusion

We found a way to determine causal connections in massive time series data. Also, we discovered such connections between the price change as an effect caused by combinations of specific cognitive distortions and sentiment patterns in online media content as well as changes of trade sell and buy volumes and imbalances between them on daily basis applied to Bitcoin cryptocurrency. That gives us hope to build a solution for reliable price prediction mechanisms usable for financial applications.

## References

1. Raheman, A., Kolonin, A., Goertzel, B., Hegyközi, G., and Ansari, I.: Architecture of Automated Crypto-Finance Agent. In: 2021 International Symposium on Knowledge, Ontology, and Theory (KNOTH), pp. 10-14, doi: 10.1109/KNOTH54462.2021.9686345 (2021).
2. Raheman, A., Kolonin, A., Ansari, I.:Adaptive Multi-strategy Market Making Agent. In: Goertzel B., Iklé M., Potapov A. (eds) Artificial General Intelligence. AGI 2021. Lecture Notes in Computer Science, vol 13154. Springer, Cham. https://doi.org/10.1007/978-3-030-93758-4_21 (2021).
3. Goertzel, B., Iklé, M., Goertzel, I., Heljakka, A.: Probabilistic Logic Networks: A Comprehensive Framework for Uncertain Inference. 1st Edition. 2nd Printing. Springer, 2008, ISBN-13: 978-0387768717, ISBN-10: 0387768718 (2008).
4. Kovalerchuk, B., Vityaev. E.: Data Mining for Financial Applications. In: Maimon O., Rokach L. (eds) Data Mining and Knowledge Discovery Handbook. Springer, Boston, MA. https://doi.org/10.1007/978-0-387-09823-4_60 (2009).
5. Mastakouri, A., Schölkopf, B., Janzing, D.: Necessary and sufficient conditions for causal feature selection in time series with latent common causes. arXiv:2005.08543 [stat.ME] (2020).
6. Deveikyte, J., Geman, H., Piccari, C., Provetti, A.: A Sentiment Analysis Approach to the Prediction of Market Volatility. arXiv:2012.05906 [q-fin.ST] (2020).
7. Arévalo, A., Nino, J., Hernandez, G., Sandoval, J.: High-Frequency Trading Strategy Based on Deep Neural Networks. In: Conference: International Conference on Intelligent Computing, DOI: 10.1007/978-3-319-42297-8_40 (2016).
8. Tsantekidis,A., Passalis, N., Tefas, A. Kanniainen, J., Gabbouj, M., Iosifidis, A.: Using Deep Learning for price prediction by exploiting stationary limit order book features. arXiv:1810.09965 [cs.LG] (2018).
9. Bollen, J., Thij, M., Breithaupt, F., Barron, A., Rutter, L., Lorenzo-Luaces, L., Scheffer, M.: Historical language records reveal a surge of cognitive distortions in recent decades. PNAS July 27, 2021 118 (30) e2102061118; https://doi.org/10.1073/pnas.2102061118 (2021).
10. Kolonin, A.: High-performance automatic categorization and attribution of inventory catalogs. In: Proceedings of All-Russia conference Knowledge Ontology Theories (KONT-2013), Novosibirsk, Russia, (2013), arXiv:2202.08965 [cs.IR] (2022).